\documentclass{birkmult}
\IfFileExists{mymtpro2.sty}%
	{\usepackage[mtpccal,mtpscr,subscriptcorrection]{mymtpro2}
		\let\tilde\wtilde
  	\let\cup\cupprod
		\let\cap\capprod
		\def\bigcup{\bigcupprod\limits}
		}%
	{\usepackage{amssymb}}  

  \newtheorem{thm}{Theorem}[section]
  \newtheorem{cor}[thm]{Corollary}
  \newtheorem{lem}[thm]{Lemma}
  
\theoremstyle{definition}
  \newtheorem{defn}[thm]{Definition}
\theoremstyle{remark}

  \newtheorem{myremarks}[thm]{Remarks}
  \newtheorem*{ex}{Example}
  \newtheorem*{exs}{Examples}
\DeclareMathOperator{\supp}{supp}
\DeclareMathOperator{\bond}{bond}
\DeclareMathOperator{\site}{site}
\DeclareMathOperator{\dist}{dist}
\DeclareMathOperator{\Aut}{Aut}
\DeclareMathOperator{\tr}{tr}
\DeclareMathOperator{\per}{per}
\DeclareMathOperator{\Id}{Id}
\DeclareMathOperator{\e}{e}
\newcommand{\BB}{\mathbb{B}}
\newcommand{\NN}{\mathbb{N}}
\newcommand{\ZZ}{\mathbb{Z}}
\newcommand{\RR}{\mathbb{R}}

\newcommand{\LL}{\mathbb{L}}
\newcommand{\EE}{\mathbb{E}}
\newcommand{\PP}{\mathbb{P}}

\newcommand{\zd}{\ZZ^{d}}
\newcommand{\bbk}{\BB_{\kappa}}

\def\DN{\Delta^{{N}}}
\def\DD{\Delta^{{D}}}
\def\DX{\Delta^{{X}}}
\def\NDT{N_{{A}}}
\def\NX{N_{{X}}}
\def\ND{N_{{D}}}
\def\NNn{N_{{N}}}

\let\emptyset\varnothing
\renewcommand{\le}{\leqslant}
\renewcommand{\ge}{\geqslant}

\numberwithin{equation}{section}

\makeatletter

\newenvironment{remarks}{\begin{myremarks}\begin{nummer}}%
    {\end{nummer}\end{myremarks}}  

\newcounter{numcount}
\newcommand{\labelnummer}{\textup{(\arabic{numcount})}}%

\newenvironment{nummer}%
{\let\curlabelspeicher\@currentlabel%
  \begin{list}{\labelnummer}{\usecounter{numcount}\leftmargin0pt%
      \topsep0.5ex\partopsep2ex\parsep0pt\itemsep0.5ex\@plus1\p@%
      \labelwidth3.5em\itemindent3.5em\labelsep1em}%
    \let\saveitem\item%
    \def\item{\saveitem%
      \def\@currentlabel{\curlabelspeicher$\,$\labelnummer}%
      \let\label\bemlabel}}%
  {\end{list}}%

\AtBeginDocument{}

\AtBeginDocument{\let\yetanotherlabel\label}
\def\bemlabel#1{\yetanotherlabel{#1}
  \def\@currentlabel{\labelnummer}
  \yetanotherlabel{#1item}}%

\def\pper{.}
\def\HarvardComma{}

\newcounter{aucount}
\newif\ifedplural
\newif\ifper\pertrue

\def\au#1#2{{#1 #2}}
\def\lau#1#2{{#1 #2}, }
\def\ed#1#2{\ifnum\theaucount=0\relax\fi{#1 #2}\addtocounter{aucount}{1}}
\def\led#1#2{\ifnum\theaucount=0\relax\edpluralfalse\else\edpluraltrue\fi{#1
    #2} (\editorname.),\setcounter{aucount}{0}}
\def\editorname{\ifedplural Eds\else Ed\fi}

\def\et{\ifnum\theaucount=1\else\HarvardComma\fi{} and\ }
\def\ti#1{\emph{#1}.\ifper\fi\pertrue}

\def\bti{\@ifnextchar[\bbti\bbbti}
\def\bbti[#1]#2{\emph{#2}, #1.}
\def\bbbti#1{\emph{#1}.}

\def\z{\@ifnextchar[\zz\zzz}
\def\zz[#1]#2#3#4#5{\perfalse\emph{#2} \textbf{#3}, \ifx
  @#5@\relax\else (#5)\fi, #3 [#1]\ifper\pper\fi\pertrue} 
\def\zzz#1#2#3#4{{#1} \textbf{#2} \ifx @#4@\relax\else
  (#4)\fi, #3\ifper\pper\fi\pertrue}

\def\pub{\@ifstar\pubstar\pubnostar}
\def\pubnostar{\@ifnextchar[\@@pubnostar\@pubnostar}
\def\@@pubnostar[#1]#2#3#4{#2, #3, #4, #1\ifper\pper\fi\pertrue}
\def\@pubnostar#1#2#3{#1, #2, #3\ifper\pper\fi\pertrue}
\def\pubstar[#1]#2#3#4{\perfalse #2, #3, #4 [#1]\pper\pertrue}

\makeatother

\newcommand{\Hmm}[1]{\leavevmode{\marginpar{\tiny%
$\hbox to 0mm{\hspace*{-0.5mm}$\leftarrow$\hss}%
\vcenter{\vrule depth 0.1mm height 0.1mm width \the\marginparwidth}%
\hbox to 0mm{\hss$\rightarrow$\hspace*{-0.5mm}}$\\\relax\raggedright #1}}}
\begin{document}
%
%
%
%
%
%
%
%
\title[Percolation Hamiltonians]{Percolation Hamiltonians}
\author[Peter M\"uller]{Peter M\"uller}

\address{%
Mathematisches Institut der Universit\"at M\"unchen\\
Theresienstr. 39\\
D-80333 M\"unchen
}

\email{mueller@lmu.de}

\thanks{To appear in:	\ed{D.}{Lenz}, \ed{F.}{Sobieczky}\et\led{W.}{Woess}
	\bti[Progress in Probability, vol. 64]{Boundaries and spectra of random walks. 
	Proceedings of the alp workshop Graz - St. Kathrein 2009} 
	\pub{Springer}{Basel}{2011}
}

\author{Peter Stollmann}
\address{Fakult\"at f\"ur Mathematik\\
TU-Chemnitz\\
D-09107 Chemnitz}
\email{peter.stollmann@mathematik.tu-chemnitz.de}


\date{\today}

\begin{abstract}
There has been quite some activity and progress concerning spectral asymptotics of random operators that are defined on percolation subgraphs of different types of graphs. In this short survey we record some of these results and explain the necessary background coming from different areas in mathematics: graph theory, group theory, probability theory and random operators.
\end{abstract}

\maketitle

\section{Preliminaries}

Here we record basic notions, mostly to fix notation. Since this survey is meant to be readable by experts from different communities, this will lead to the effect that many readers might find parts of the material in this section pretty trivial -- never mind.

%
%

\subsection{Graphs}
A graph is a pair
$G=(V,E)$ consisting of a countable set of \emph{vertices} $V$ together with a set $E$ of
\emph{edges}. Since we consider undirected graphs without loops, edges can and will be regarded as subsets $e=\{ x,y\}\subseteq V$. In this case we say that $e$ is an edge between $x$ and $y$, respectively adjacent to $x$ and $y$. Sometimes  we write $x \sim y$ to indicate that $\{ x,y\}\in E$. The \emph{degree}, the number of edges adjacent to $x$, is denoted by
$$
\deg_G:=\deg:V\to\NN_0, \deg(x):=\#\{y\in V\mid x\sim y\} .
$$
A graph with constant degree equal to $k$ is called a $k$-regular graph.

A \emph{path} is a finite family $\gamma:=(e_1,e_2,...,e_n)$ of consecutive edges,
 i.e., such that $e_k\cap e_{k+1}\not =\emptyset$; the set of points visited by $\gamma$ is denoted by $\gamma^*:=e_1\cup ...\cup e_n$. This gives a natural notion of \emph{clusters} or \emph{connected components} as well as a natural distance in the following way. If $x$ is a vertex, then $C_x$, the \emph{cluster containing} $x$, is the set of all vertices $y$, for which there is a path  $\gamma$ \emph{joining} $x$ and $y$, i.e., so that $x,y\in\gamma^*$. The length of a shortest path joining $x$ and $y$ is called the \emph{distance} $\dist(x,y)$. With the convention $\inf\emptyset :=\infty$ it is defined on all of $V$, its restriction to any cluster induces a metric.

A \emph{subgraph} $G'=(V',E')$ of $G$ is given by a subset $V'\subseteq V$ and a subset $E'\subseteq E$. The \emph{subgraph $G'=(V',E')$ induced by} $V'$ has the edge set $E'=\{ e\in E\mid e\subseteq V'\}$.

A one-to-one mapping $\Phi:V\to V$ is called an \textit{automorphism} of the graph $G=(V,E)$ if $\{ x,y\}\in E$ if and only if $\{ \Phi(x),\Phi(y)\}\in E$. The set of all automorphisms $\Aut(G)$ is a group, when endowed with the composition of automorphisms as group operation. An \textit{action} of a group $\Gamma$ on $G$ is a group homomorphism $j:\Gamma\to \Aut(G)$, and we write $\gamma x:=(j(\gamma))(x)$ for $\gamma\in\Gamma, x\in V$. An action is called \textit{free}, if $\gamma x=x$ only happens for the neutral element $\gamma=e$ of $\Gamma$. A group action is called \textit{transitive}, if the \emph{orbit} $\Gamma x:= \{\gamma x\mid \gamma \in\Gamma\}$ of $x$ equals $V$ for some (and hence every) vertex $x\in V$. Note that in this case $G$ looks the same everywhere. 

\begin{ex}
A prototypical example is given by the \emph{$d$-dimensional integer lattice graph} $\LL^d$ with vertex set $\zd$ and edge set given by all unordered pairs of vertices with Euclidean distance one. Clearly, the additive group $\zd$ acts transitively and freely on $\LL^d$ by translations.
\end{ex}

For any group action, due to the group structure of $\Gamma$, it is clear that two orbits $\Gamma x\not= \Gamma y$ must be disjoint. If there are only a finite number of different orbits under the action of $\Gamma$, the action is called \textit{quasi-transitive}, in which case there are only finitely many different ways in what the graph can look like locally.
For quasi-transitive actions, there are finite minimal subsets $\mathcal{F}$ of $V$ so that 
\begin{equation}
	\label{quasitr}
 \bigcup_{x\in\mathcal{F}}\Gamma x=V.
\end{equation}
These are called \textit{fundamental domains}.

%
%

\subsection{The adjacency operator and Laplacians}\label{laplacians}

The \emph{adjacency operator} of a given graph $G=(V,E)$ acts on the Hilbert space $\ell^{2}(V)$ of complex-valued, square-summable functions on $V$ and is given by
$$A:=A_G:\ell^2(V)\to \ell^2(V), Af(x):=\sum_{y\sim x}f(y)
\qquad \text{for} \; f\in \ell^2(V),\; x\in V.
$$
We will assume throughout that the degree $\deg$ is a bounded function on $V$, and so $A$ is a bounded linear operator. The (combinatorial or graph) \emph{Laplacian} is defined as
$$\Delta:=\Delta_G:\ell^2(V)\to \ell^2(V), \Delta f(x):=\sum_{y\sim x}[f(x)-f(y)]
\qquad \text{for} \; f\in \ell^2(V),\; x\in V
$$
so that $\Delta_G = D_G-A_G$, where $D:=D_G$ denotes the bounded multiplication operator with $\deg$. Signs are a notorious issue here: note that (contrary to the convention in most of the second author's papers) there is no minus sign in front of the triangle.

For a subgraph $G'=(V',E')$ of a given graph, certain variants of $\Delta_{G'}$ are often considered:
The \emph{Neumann Laplacian} is just
$\Delta_{G'}^N := \Delta_{G'}$, meaning that the ambient larger graph plays no role at all. The \emph{Dirichlet Laplacian} $\Delta_{G'}^D$ (the notation agrees with that of \cite{KiMu04,AntunovicV-08,AntunovicV-09,MuellerS-07}) penalises boundary vertices of $G'$ in $G$, that is vertices with a lower degree in $G'$ than in $G$:
$$
\Delta_{G'}^D:= 2(D_{G} -D_{G'}) + \Delta_{G'}^N = 2D_G - D_{G'}-A_{G'}:\ell^2(V')\to\ell^2(V') .
$$
A third variant is called pseudo-Dirichlet Laplacian in \cite{KiMu04,MuellerS-07}; here we use the notation from \cite{AntunovicV-08,AntunovicV-09}, where it is named
adjacency Laplacian:
$$
	\Delta_{G'}^A:=D_{G} -D_{G'} + \Delta_{G'}^N = D_G-A_{G'}:\ell^2(V')\to\ell^2(V') .
$$
 The motivation and origin for the terminology of the different  boundary conditions are discussed in \cite{KiMu04} -- together with some
  basic properties of these operators. Most importantly, they are ordered in the sense of quadratic forms
\begin{equation}
\label{order}
	0 \le \Delta_{G'}^N \le \Delta_{G'}^{A} \le \Delta_{G'}^D \le 2 D_{G} \le 2 \|\!\deg_{G}\!\|_{\infty} \Id
\end{equation}  
on $\ell^2(V')$. Here, $\Id$ stands for the identity operator. We recall that for bounded operators on a Hilbert space $\mathcal{H}$, the partial ordering $A \le B$ means $\langle \psi, (B-A) \psi\rangle \ge 0$ for all $\psi\in\mathcal{H}$, where the brackets denote the scalar product on $\mathcal{H}$. Thus the spectrum of each Laplacian $\Delta_{G'}^X$, $X\in \{N, A, D\}$, is confined according to $\mathrm{spec} (\Delta_{G'}^X) \subseteq \big[0, 2\|\deg_{G}\|_{\infty}\big]$. The names Dirichlet and Neumann are chosen in reminiscence of the different boundary conditions of Laplacians on open subsets of Euclidean space. In fact one can easily check that for disjoint subgraphs $G_1, G_2\subset G$,
$$
\Delta^N_{G_1} \oplus \Delta^N_{G_2}\le \Delta^N_{G_1\cup G_2}\le
                            \Delta^D_{G_1\cup G_2}\le
                     \Delta^D_{G_1} \oplus\Delta^D_{G_2} .
$$
The adjacency Laplacian does not possess such a monotonicity.

On bipartite graphs, such as the lattice graph $\LL^d$, the different Laplacians are related to each other by a special unitary transformation on $\ell^2(V)$. We recall that a graph is \emph{bipartite} if its vertex set can be decomposed into two disjoint subsets $V_{\pm}$ so that no edge joins two vertices within the same subset. Define a unitary involution $U=U^*=U^{-1}$ on $\ell^{2}(V)$ by
$(Uf)(x):= \pm f(x)$ for $x\in V_{\pm}$. Clearly, we have $U^*DU = D$ and $U^*AU = - A$. The latter holds because of
\begin{equation*}
 \bigl( A(Uf) \bigr)(x) = \sum_{y\sim x} (Uf)(y) = \sum_{y\sim x} \mp f(y)	 
 	= - \bigl( U (Af) \bigr)(x)
\end{equation*}
for every $x \in V_{\pm}$. In particular, for any subgraph $G'$ of a $k$-regular bipartite graph 
$G$ we get
\begin{equation} 
\label{lap-symm}
\begin{split}
 \Delta^A_{G'}& =  2k\Id - U^* \Delta^A_{G'}U \\
\Delta^N_{G'}& =  2k\Id - U^* \Delta^D_{G'}U \\
\Delta^D_{G'}& =  2k\Id - U^* \Delta^N_{G'}U .
\end{split}
\end{equation}
Consequently, spectral properties of the different Laplacians at zero -- the smallest possible spectral value as allowed by \eqref{order} -- can be translated into spectral properties (of another Laplacian) at $2k$.

%
%

\subsection{Amenable groups and their Cayley graphs}

Here we record several basic notions and results that will be used later on; we largely follow \cite{AntunovicV-08}.

Let $\Gamma$ be a finitely generated  group and $S\subset \Gamma$ a symmetric (i.e.\ $S^{-1}\subseteq S$) finite set of generators that does not contain the identity element $e$ of $\Gamma$. The \emph{Cayley graph} $G=G(\Gamma,S)$ has $\Gamma$ as a vertex set and an edge connecting $x,y\in \Gamma$ provided $xy^{-1}\in S$. By symmetry of $S$ we get an undirected graph in this fashion, and $G$ is $|S|$-regular. Moreover, it is clear that $\Gamma$ acts transitively and freely on $G$ by left multiplication.

\begin{exs}
 \begin{nummer}
  \item The $d$-dimensional integer lattice graph $\LL^d$ is the Cayley graph of the group $\ZZ^d$ (written additively, of course) with the set of generators $S=\{ e_j, -e_j\mid j=1,...,d\}$ with $e_j$ the unit vector in direction $j$.
   \item Changing the set of generators to $S':=S\cup \{ \pm e_j \pm e_k\mid 1 \le j < k \le d \}$ gives additional diagonal edges; see Figure 1 for an illustration in $d=2$.

\begin{figure}
	\includegraphics[width=0.35\textwidth]{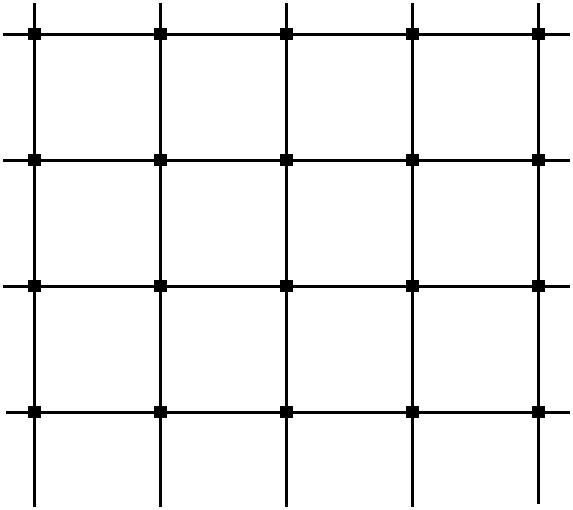}\hspace*{2.5cm}\includegraphics[width=0.35\textwidth]{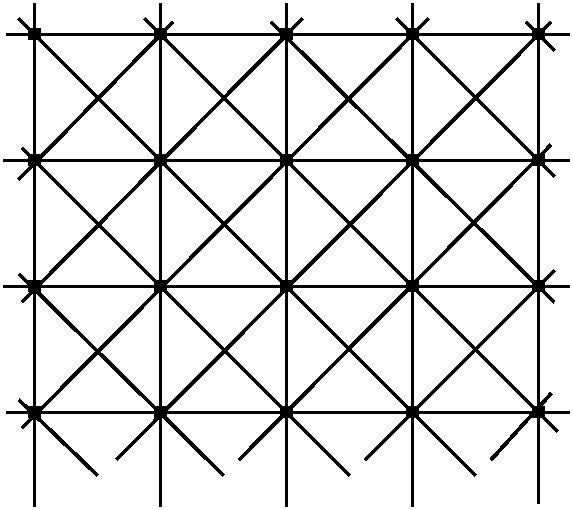}
	\caption{Two Cayley graphs of $\ZZ^d$.}
	\label{fig:1}
\end{figure}

   \item The Cayley graph of the free group with $n \in\NN \setminus \{1\}$ generators $g_1,...,g_n$ can be formed with $S=\{  g_1,...,g_n, g_1^{-1},...,g_n^{-1}\}$; it is a $2n$-regular rooted infinite tree. More generally, a $(\kappa +1)$-regular rooted infinite tree, $\kappa \in \NN \setminus  \{1\}$, is also called \textit{Bethe lattice} $\mathbb{B}_{\kappa}$, honouring Bethe \cite{Bethe-35} who introduced them as a popular model of statistical physics. Every vertex other than the root $e$ in $\mathbb{B}_{\kappa}$ possesses one edge leading ``towards'' the root and $\kappa$ ``outgoing'' edges, see Figure~2 for an illustration for $n=2$, respectively $\kappa=3$.
\end{nummer}
\end{exs}

\suppressfloats[t]

\begin{figure}[t]
	\includegraphics[width=0.35\textwidth]{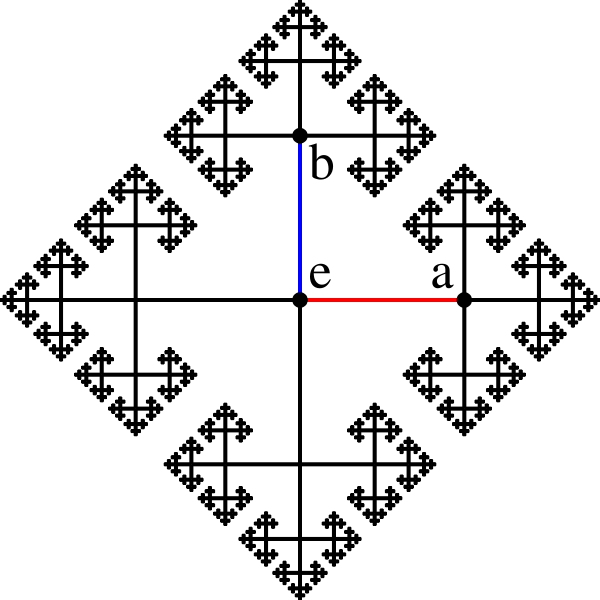}
	\caption{Bethe lattice $\mathbb{B}_{3}$, the Cayley graph of the free group with $n=2$ generators $a,b$.}
	\label{fig:2}
\end{figure}

Due to fundamental theorems of Bass \cite{Bass-72}, Gromov \cite{Gromov-81} and van den Dries and Wilkie \cite{vandenDriesW-84}, the volume, i.e.\ the number of elements, of the  ball  $B(n)$ consisting of all those vertices that are at distance at most $n$ from the identity $e$,
\begin{equation}
\label{volume}
 V(n):= | B(n)| := \# \big\{ x \in \Gamma \mid \dist_{G(\Gamma,S)}(x,e) \le n \big\},
\end{equation}
has an asymptotic behaviour that obeys one of the following alternatives:

\begin{thm}\label{growth}
 Let $G=G(\Gamma,S)$ be the Cayley graph of a finitely generated group. Then exactly one of the following is true:
\begin{itemize}
 \item [(a)] $G$ has polynomial growth, i.e., $V(n)\sim n^d$ for some $d\in\NN$.
  \item [(b)] $G$ has superpolynomial growth, i.e., for all $d\in\NN$ and $b\in\RR$ there are only finitely many $n\in\NN$ so that $V(n)\le b n^d$.
\end{itemize}
The growth behavior, in particular the exponent $d$, is independent of the chosen set $S$ of generators.
\end{thm}

There is another issue of importance to us, \textit{amenability}. A definition in line with our subject matter here goes as follows:
\begin{defn}
 A discrete group $\Gamma$ is called \textit{amenable}, if there is a \textit{F\o{}lner sequence}, i.e., a sequence $(F_n)_{n\in \NN}$ of finite subsets which exhausts $\Gamma$ with the property that for every finite $F\subset \Gamma$:
$$
\frac{|(F\cdot F_n)\triangle F_n|}{| F_n |}\to 0 \qquad \text{for \quad}n\to\infty ,
$$
where $A \triangle B:=(A\setminus B)\cup (B\setminus A)$ denotes the symmetric difference of two sets $A$ and $B$.
\end{defn}
There is quite a number of different equivalent characterisations of amenability. The notion goes back to John von Neumann \cite{vonNeumann-29}. In its original form he required the existence of a \textit{mean} on $\ell^\infty (\Gamma)$, i.e., a positive, normed, $\Gamma$-invariant functional.
\begin{remarks}
 \item The defining property of a F\o{}lner sequence is that the volume of the boundary of $F_n$ becomes small with respect to the volume of $F_n$ itself as $n\to\infty$. Boundary as a topological term is of no use here; instead, thinking of the associated Cayley graph, $F\cdot F_n$ can be thought of as a neighborhood around $F_n$ (at least for $F$ containing the identity) and so $ |(F\cdot F_n)\triangle F_n|$ represents the volume of a boundary layer around $F_n$. Thinking of $F$ as the ball $B(r)$ makes this picture quite suggestive.
\item Discrete groups of subexponential growth are amenable.
 \item The \textit{lamplighter groups} (see below) are amenable but not of subexponential  growth. Consequently, growth does not determine amenability.
 \item The standard example of a nonamenable group is the free group on two generators.
\end{remarks}
 
\noindent 
Let us end this subsection with the example we already referred to above:
\begin{ex}
 Fix $m\in \NN, m\ge 2$. The \textit{wreath product} $\ZZ_m\wr \ZZ$ is the set
$$
\ZZ_m\wr \ZZ :=\{ (\varphi,x)\mid \varphi :\ZZ\to \ZZ_m, \supp\varphi \mbox{  finite  }, x\in\ZZ \},$$
$$
(\varphi_1,x_1)*(\varphi_2,x_2):= (\varphi_1+\varphi_2(\cdot -x_1), x_1+x_2)
$$
and is called the lamplighter group. It is amenable, see \cite{AntunovicV-09}.
\end{ex}

%
%

\section{Spectral asymptotics of percolation graphs}
This section contains the heart of the matter of the present survey. After introducing percolation, we begin discussing the relevant properties of the random operators associated with percolation subgraphs. The central notion is the \textit{integrated density of states}, a real-valued function. We then explain a number of results on the asymptotic behaviour of this function and how methods from analysis, geometry of groups, graph theory and probability are used to derive these results.

%
%

\subsection{Percolation}

Percolation is a probabilistic concept with a wide range of applications, usually related to some notion of \textit{conductivity} or \textit{connectedness}.
Its importance in (statistical) physics lies in the fact that, despite its simplicity, percolation yet exposes a phase transition. The mathematical origin of percolation can be traced back to a question of Broadbent that was taken up in two fundamental papers by Broadbent and Hammersley in 1957 \cite{BroadbentH-57,Hammersley-57}. Percolation theory still has an impressive list of easy-to-state open problems to offer, some with well established numerical data and conjectures based on physical reasoning. We refer to \cite{Gri99,Kesten-82} for standard references concerning the mathematics, as well as Kesten's recent article in the Notices of the AMS \cite{Kesten-06}.

Mathematically speaking, and presented in accordance with our subject matter here, percolation theory deals with random subgraphs of a given graph $G=(V,E)$ that is assumed to be infinite and connected.  A good and important example is the $d$-dimensional lattice graph $\LL^d$, the particular case $d=1$ being very special, however. There are two different but related random procedures to delete edges and vertices from $G$, called \textit{site percolation} and \textit{bond percolation}. In both cases, everything will depend upon one parameter $p\in [0,1]$ that gives the probability of keeping vertices or edges, respectively. 

Let us start to describe \emph{site percolation}. We consider the infinite product
$$
\Omega:=\Omega^{\site}:=\{0,1\}^V, \qquad
\PP_p:= \bigotimes_{x\in V} \big(p\cdot \delta_1+(1-p)\cdot \delta_0\big),
$$
as probability space with elementary events $\omega := (\omega_{x})_{x\in V}$, $\omega_{x}\in \{0,1\}$, and a product Bernoulli measure $\PP_{p}$ that formalizes the following random procedure. Independently for all vertices (also called \emph{sites} in this context) of $V$, we delete the vertex $x$ from the graph with probability $1-p$, along with all edges adjacent to $x$. This corresponds to the event $\omega_x=0$, and we call the site $x$ \emph{closed}. On the other hand, we keep the vertex $x$ and its adjacent edges in the graph with probability $p$. This corresponds to the event $\omega_x=1$, in which case we speak of an \textit{open} site. Every possible realisation or configuration is given by exactly one element $\omega=(\omega_x)_{x\in V}\in\Omega$, and the measure $\PP_p$ above governs the statistics according to the rule we just mentioned. Note that we omit the superscript in the notation of the product measure. The graph we just described is illustrated in Figure 3 and formally defined by $G_\omega=(V_\omega,E_\omega)$, where 
$$
V_\omega:=\{ x\in V\mid \omega_x=1\}, \qquad
E_\omega:=\{ e\in E\mid e\subseteq V_\omega\} ,
$$
i.e.\ the subgraph of $G$ induced by $V_{\omega}$.
Note that for  $p=0$ the graph $G_{\omega}$ is empty with probability $1$ and for $p=1$ we get $G_\omega =G$ with probability 1.

\begin{figure}[t]\label{fig:3}
 \includegraphics[width=0.4\textwidth]{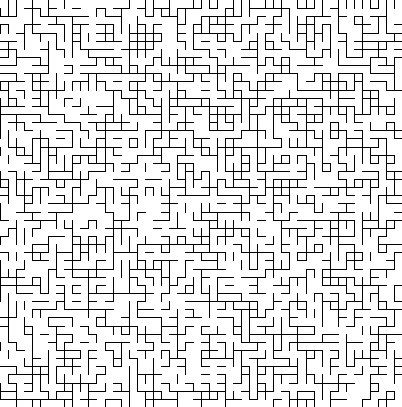}
\caption{Part of a realisation $G_{\omega}$ for bond percolation on $\LL^2$ for $p=\frac12$.}
\end{figure}

The second variant, \emph{bond percolation}, works quite similarly:
$$
\Omega:=\Omega^{\bond}:=\{0,1\}^E,\qquad
\PP_p:= \bigotimes_{x\in E}(p\cdot \delta_1+(1-p)\cdot \delta_0),
$$
leading to the subgraph $G_\omega=(V_\omega,E_\omega)$ with
$$
V_\omega:=V, \qquad E_\omega:=\{ e\in E\mid \omega_e=1 \} .
$$
It amounts to deleting edges (also called \emph{bonds} in this context) with probability $1-p$, independently of each other. The choice $V_{\omega} =V$ is merely a convention. Other authors keep only those vertices that are adjacent to some edge.
 
In both site and bond percolation, the issue is the connectedness of the so-obtained random subgraphs. Note that the realisations $G_\omega$ themselves do not depend upon $p$, while assertions concerning the probability of certain events or the stochastic expectation of random variables constructed from the subgraphs surely do. A typical question is whether the cluster $C_x$ that contains vertex $x\in V$ is finite in the subgraph $G_{\omega}$ for $\PP_{p}$-almost all $\omega\in \Omega$ or whether it is infinite with non-zero probability. In the latter case one says that percolation occurs. 

Let us assume from now on that $G$ is quasi-transitive, so that the above question will have an answer that is independent of $x$. The \textit{percolation threshold} or \textit{critical probability} is then defined as
$$
p_H:=\sup\big\{ p\in [0,1] \big|\; \PP_p[ |C_x|=\infty ]=0\big\} .
$$
It is independent of $x$ since, globally, $G$ looks the same everywhere, cf.\ \eqref{quasitr}, and $\PP_{p}$ is a product measure consisting of identical factors. A related critical value is given by
$$
p_T:=\sup\big\{ p\in [0,1] \big| \; \EE_p [|C_x|] < \infty \big\} ,
$$
and it is clear that $p_T\le p_H$. Here, $\EE_{p}$ stands for the expectation on the probability space $(\Omega,\PP_p)$. The equality of these two critical values is often dubbed \textit{sharpness of the phase transition}, and we write $p_{c} := p_{H} = p_{T}$ in this case for \emph{the} critical probability. Clearly, sharpness of the transition is a desirable property, as both $p_{H}$ and $p_{T}$ represent two equally reasonable ways to distinguish a phase with $\PP_{p}$-almost surely only finite clusters, 
the \textit{subcritical} or \emph{non-percolating} phase, from a phase where there exists an infinite cluster with probability one, the \textit{supercritical} or  \emph{percolating} phase. Apart from that, sharpness of the phase transition has been used as an important ingredient in the proof of Kesten's classical result that $p_c=\frac12$ for bond percolation on the 2-dimensional integer lattice $\LL^2$. Together with estimates known for $p<p_T$, it gives that the expectation of the cluster size decays exponentially, i.e.,
$$
\PP_p\{ |C_x|=n \}\le e^{-\alpha_pn} , \qquad n\in\NN,
$$
with some constant $\alpha_{p} >0$ for all $p<p_c$. This fact is also heavily used in some proofs of Lifshits tails for percolation subgraphs, see below. Fundamental papers that settle sharpness of the phase transition for lattices and certain quasi-transitive percolation models are
\cite{AizenmanB-87,Men'shikov-86,Men'shikovMS-86}. Recent results valid for all quasi-transitive graphs can be found in \cite{AntunovicV-08b} together with a discussion of the generality of earlier literature. 

\begin{thm}  \textup{\!\! (\cite{AntunovicV-08b}, Theorem 2, Theorem 3)\;}\label{sharp}
 For every quasi-transitive graph
$$p_T=p_H =:p_{c} ,$$
and for every $p<p_c$ there exists a constant $\alpha_p >0$ so that
$$
\PP_p\{ |C_x|\ge n\} \le e^{-\alpha_p n}\mbox{   for all   }
x\in V, n\in\NN .
$$
\end{thm}

It is expected that sharpness of the phase transition also holds for percolation on more general well-behaved graphs even without quasi-transitivity. 
The celebrated Penrose tiling gives rise to such a graph without quasi-transitivity but some form of aperiodic order. A result analogous to Thm.\ \ref{sharp} was proven for the Penrose tiling in \cite{Hof-98}. The general case of graphs with aperiodic order has not yet been settled. We refer to \cite{MuellerR-07} for partial results in this direction.

%
%

\subsection{The integrated density of states}\label{ids}

The study of the random family $(\Delta_{G_\omega})_{\omega\in\Omega}$ of Laplacians on percolation graphs was proposed by de Gennes \cite{deGennesLM-59a,deGennesLM-59b} and often runs under the header quantum percolation in physics. In this paper we focus on the \emph{integrated density of states} (IDS), also called \textit{spectral distribution function}, of this family of operators. 

In general, the IDS is the distribution function of a (not necessarily finite) measure on $\RR$ that is meant to describe the density of spectral values of a given self-adjoint operator. In the cases of interest to us here, the underlying Hilbert space is $\ell^2(V)$, with $V$ being the countable vertex set of some graph. In this situation the IDS is even the distribution function of a probability measure on $\RR$, as we shall see. Before giving the rigorous definition that applies in this setting, let us first start with a discussion at a heuristic level.  For elliptic operators acting on functions on some infinite configuration space $V$ with a periodic geometric structure, one typically does not have eigenvalues, but rather continuous spectrum. However, the restrictions of these operators to compact subsets $K$ of configuration space $V$ (more precisely to $\ell^2(K)$, actually) come with discrete spectrum.
Therefore, one can count eigenvalues, including their multiplicities. The idea of the IDS is to calculate the number of eigenvalues per unit volume for an increasing sequence $K_n$ of compact subsets and take the limit. For this procedure to make sense, the operator has to be homogenous, at least on a statistical level. Two situations are typical: Firstly, a periodic operator, quite often the Laplacian of a periodic geometry. And, secondly, an ergodic (statistically homogenous) random family of operators, in which case the above mentioned limit will exist with probability one.

Let $H$ be a self-adjoint operator in $\ell^2(V)$. An intuitive ansatz for the definition of the IDS might be $N: \RR \to  [0,1]$,
\begin{equation}
	\label{ids-0}
  E \mapsto  N(E) := \displaystyle \lim_{n\to\infty}\frac{\tr\big[1_{F_n}1_{]-\infty,E]}(H)\big]}{| F_n|} 
= \lim_{n\to\infty} 
\frac{\sum_{x\in F_{n}} \langle \delta_{x}, 1_{]-\infty,E]}(H)\delta_{x}\rangle}{|F_{n}|} \,  ,
\end{equation}
where $(F_n)_{n\in\NN}$ is an appropriate sequence of finite sets exhausting $V$. Before we go on, let us add some remarks on our notation in \eqref{ids-0}. In general, we write $1_{A}$ for the indicator function of some set $A$. Above, $1_{F_{n}}$ is to be interpreted as the multiplication operator corresponding to the indicator function $1_{F_{n}}$. In view of the functional calculus for self-adjoint operators we write $1_{B}(H)$ for the spectral projection of $H$ associated to some Borel set $B \subseteq \RR$. Finally, $\tr$ stands for the trace on $\ell^2(V)$ and $\delta_{x} \in \ell^2(V)$ for the canonical basis vector that is one at vertex $x$ and zero everywhere else.  

As was already mentioned, a certain homogeneity property is necessary in order for the limit in \eqref{ids-0} to exist. A careful choice of the exhausting sequence is necessary, too. For amenable groups tempered F\o{}lner sequences will do the job, as is ensured by a general ergodic theorem of Lindenstrauss \cite{Lindenstrauss-01}. We refer to \cite{LenzV-08, MuellerR-07, MuellerR-10} for more details in the present context and sum up the main points in the following definition and the subsequent results.

\begin{defn}
\label{ids-def}
 Let $G$ be a graph and let $\Gamma$ be an infinite group that acts
quasi-transitively on $G$. We fix a fundamental domain $\mathcal{F}$. For $E\in\RR$ we define
\begin{equation}
	\label{ids-per}
 	N_{\per} (E):=\frac{1}{|\mathcal{F}|}\tr \big[ 1_\mathcal{F}1_{]-\infty,E]}(\Delta_G)\big] 
\end{equation}
to be the IDS of the full graph. Secondly, the expression
\begin{equation}
	\label{ids-random}
 	N_{X}(E) := N^{(p)}_{X}(E):=\frac{1}{|\mathcal{F}|}\; \EE_{p}\left\{\tr\left[ 1_\mathcal{F}1_{]-\infty,E]}(\Delta^X_{G_		\omega})\right]\right\}	
\end{equation}
is the IDS of the Laplacians on random percolation subgraphs, where $X\in\{ N,A,D\}$ stands for one of the possible boundary conditions discussed in Subsection \ref{laplacians}.
\end{defn}
\begin{remarks}
  \item We could have chosen a more general probability measure than $\PP_p$, as long as it is invariant under $\Gamma$.
  \item Usually, we will omit the superscript $p$ and write simply $N_{X}$ for the quantity in \eqref{ids-random}.
  \item  Note that $N_{\per} =N^{(1)}_{X}$ for any $X\in\{ N,A,D\}$.
  \item Note also that $N_{X}$ is not defined in terms of a single operator $\Delta^X_{G_\omega}$, but rather using the whole family $(\Delta^X_{G_\omega})_{\omega\in \Omega}$; see also the subsequent result for a clarification.
\end{remarks}
The next theorem establishes the connection between the  heuristic picture displayed in \eqref{ids-0} and the preceding definition. The point here is the generality of the group involved. In the more conventional setting of random operators on Euclidean space $\RR^d$ (with the group action of $\ZZ^d$), the equation is the celebrated \emph{Pastur-Shubin trace formula}.

\begin{thm}
\textup{\!(\cite{LenzV-08}, Theorem 2.4)\;}
\label{self-av}
 Let $G$ be a graph and let $\Gamma$ be an infinite group that acts
quasi-transitively on $G$. Then there is a sequence $(F_n)_{n\in\NN}$ of finite subsets of $V$ so that
\begin{equation}
\label{mac-limit}
N_X(E)=\lim_{n\to\infty} \frac{1}{|{F_n}|}\tr\left[ 1_{]-\infty,E]}(1_{F_n}\Delta^X_{G_\omega}1_{F_n})\right] ,
\end{equation}
uniformly in $E\in\RR$ for $\PP_{p}$-a.e. $\omega\in\Omega$.
\end{thm}

\begin{remarks}
\item 
	We refer to \cite{DodziukLMSY-03,KiMu04,LeMuVe08,MathaiY-02,Ves04}
	for further predecessors of the latter theorem.
\item
	The inequalities in \eqref{order} imply
	$$
		\ND \le N_{A} \le \NNn.
	$$
\item
A comprehensive theory of the IDS in the (more conventional) set-up of random Schr\"odinger operators can be found in the monographs \cite{CarmonaL-90,PasturF-92,Sto01}; see also the surveys \cite{Kirsch-85,KirschM-07,Veselic-08} and the references therein. 
\end{remarks}
 
Interestingly, the IDS links quite a number of different areas in mathematics:
We started with an elementary operator theoretic point of view. If we rephrase the basic existence problem in the way that we regard the counting of eigenvalues as evaluating the trace of the corresponding eigenprojection, we arrive at the question, whether appropriate traces exist on certain operator algebras. Typically, the operators we have in mind are intimately
linked to some geometry, so that quantities derived from the IDS play an important role in geometric analysis.  An important example is the \emph{Novikov-Shubin invariant} of order zero, which equals the \emph{van Hove exponent} in the mathematical physics language and will be discussed in our setting further below; see \cite{NovikovS-86,GromovS-91}
and the Oberwolfach report \cite{DodziukLPSV-06}.
Another wellknown principle provides a link to stochastic processes and random walks: The Laplace transform of $N_N$ is the return probability of a continuous time random walk on the graph; details geared towards the applications we have in mind can be found in \cite{MuellerS-07}. 

The original motivation and the name IDS come from physics. The Laplacians we consider show up as energy operators for a quantum-mechanical particle which undergoes a free motion on the vertices of the graph. If $v,v' \in V$ are connected by an edge, the particle can ``hop'' directly from $v$ to $v'$ or vice versa. In this way, the spectrum of the Laplacian appears as the set of possible energy values the particle may attain, hence the name IDS for the quantities in Def.~\ref{ids-def}. In the percolation case, the motion is interpreted to be a quantum mechanical motion of a particle in a random environment. Thm.~\ref{self-av} is interpreted as the \emph{self-averaging} of the IDS for a family of random ergodic operators: for $\PP$-a.e. realisation $\omega$ of the environment, the normalised finite-volume eigenvalue counting function converges to a non-random quantity. In particular, if one had taken an expectation on the r.h.s.\ of \eqref{mac-limit}, one would have ended up with the very same expression in the macroscopic limit. 

The IDS is one of the simplest, but nonetheless physically important spectral characteristics of the operators we consider. It encodes all thermostatic properties of a corresponding gas of non-interacting particles. As an example we mention a systems of electrons in a solid, where this is a reasonable approximation in many situations. Besides, the IDS enters transport coefficients such as the electric conductivity and determines the ionisation properties of atoms and molecules. For this reason, the IDS (more precisely, its derivative with respect to $E$, the \emph{density of states}) is a widely studied quantity in physics.

%
%

\subsection{The integer lattice}

In this subsection we are concerned with the asymptotics at spectral edges of the IDS of the family of Laplacians $(\DX_{G_{\omega}})_{\omega\in\Omega}$ on bond-percolation subgraphs of the $d$-dimensional integer lattice graph $\LL^d$ (or bond percolation on $\zd$, for short).

The spectral edges of these Laplacians turn out to be $0$ and $4d$. In fact, standard arguments \cite{KiMu04}, which are based on ergodicity w.r.t.\ $\zd$-translations, yield that even the whole spectrum equals almost surely the one of the Laplacian $\Delta_{\LL^d}$ on the full lattice 
$$
	\mathrm{spec} (\DX_{G_{\omega}}) = [0,4d] \qquad \text{for $\PP_{p}$-almost every $\omega\in\Omega$}, 
$$
any $p\in ]0,1]$ and $X \in \{N,A,D\}$. Thus, the left-most and right-most inequality in \eqref{order} are sharp in this case. Since the lattice $\LL^d$ is bipartite, it follows from \eqref{lap-symm} with $k=2d$ that the different Laplacians are related to each
  other by a unitary involution, which implies the symmetries
  \begin{equation}
    \label{bcrel}
    \begin{split}
      \NDT (E) &= 1 - \lim_{\varepsilon\uparrow 4d -E} \NDT(\varepsilon)\,,
      \\
      N_{{D(N)}}(E) &= 1 - \lim_{\varepsilon\uparrow 4d -E}
      N_{{N(D)}} (\varepsilon)
    \end{split}
  \end{equation}
  for their integrated densities of states for all $E\in [0,4d]$. The
  limits on the right-hand sides of \eqref{bcrel} ensure that the discontinuity
  points of $\NX$ are approached from the correct side.

As before we write $p_{c}\equiv p_{c}(d)$ for the unique critical probability of the
bond-percolation transition in $\zd$. We recall from \cite{Gri99} that $p_{c}=1$ for $d=1$, otherwise
$p_{c} \in]0,1[$.  Let us first think about what to expect. At least for small $p$, the random graph $G_{\omega}$ is decomposed into relatively small pieces, due to Theorem \ref{sharp} above. This means that there cannot be many small eigenvalues as the size of the components limits the existence of low lying eigenvalues. Consequently, the eigenvalue-counting function for small $E$ must  be small.
It turns out that the IDS vanishes even exponentially fast. This striking behaviour is called \textit{Lifshits tail}, to honour Lifshits' fundamental contributions to solid state physics of disordered systems \cite{Lifshitz-63,Lifshitz-64,Lifshitz-65}. In fact, Lifshits tails continue to show up in the percolating phase for the adjacency and the Dirichlet Laplacian at the lower spectral edge. This follows from a \emph{large-deviation principle}.  

\smallskip

\begin{thm}
	\textup{\!(\cite{MuellerS-07}, Theorem 2.5)\;}
  \label{lifshits}
  Assume $d\in\mathbb{N}$ and $p \in ]0,1[$. Then the integrated
  density of states $\NX$ of the Laplacians $(\DX_{G_{\omega}})_{\omega\in\Omega}$ on bond-percolation graphs
  in $\zd$ exhibits a \emph{Lifshits tail} at the lower
  spectral edge 
  \begin{equation}
    \label{lowerlif}
    \lim_{E\downarrow 0}\;\frac{\ln | \ln \NX(E)|}{\ln E} =
    -\; \frac{d}{2}  \qquad \text{for} \quad
    {X} \in \{{{A}}, {D}\}
  \end{equation}
  and at the upper spectral edge
  \begin{equation}
    \label{upperlif}
    \lim_{E \uparrow 4d}\;\frac{\ln |\ln [1 - \NX(E)]|}{\ln (4d -E)} =  
    -\; \frac{d}{2}  \qquad \text{for} \quad
    {X} \in \{{N}, {{A}}\} \,.
  \end{equation}
\end{thm}

Actually, slightly stronger statements without logarithms are proven in \cite{MuellerS-07}, see the next lemma. Together with the symmetries \eqref{bcrel}, these bounds will imply the above theorem.

\begin{lem}
	\textup{\!(\cite{MuellerS-07}, Lemma 3.1)\;}
  \label{liflemma}
  For every $d\in\mathbb{N}$ and every $p\in]0,1[$ there exist constants
  $\varepsilon_{{D}}$, $\alpha_{u}$, $\alpha_{l} \in ]0,\infty[$ such
  that 
  \begin{equation}
    \label{lifbounds}
    \exp\{ -\alpha_{l} E^{-d/2}\} \le \ND(E) \le \NDT(E) \le  \exp\{
    -\alpha_{u} E^{-d/2}\} 
  \end{equation}
  holds for all $E\in ]0, \varepsilon_{{D}}[$.
\end{lem}

\begin{remarks}
\item In the non-percolating phase, $p\in]0,p_{c}[$, the content of
  Theorem~\ref{lifshits} has already been known from \cite{KiMu04}, where it is proved by a different
  method. The method of \cite{KiMu04}, however, does not seem to extend to the
  critical point or the percolating phase, $p \in]p_{c},1[$.
\item 
  The Lifshits asymptotics of Theorem~\ref{lifshits} are determined by those
  parts of the percolation graphs which contain large, fully-connected cubes.
  This also explains why the spatial dimension enters the Lifshits exponent
  $d/2$.
\item \label{bikorem}
  We expect that \eqref{lowerlif} can be refined in the adjacency case ${X} =
  {{A}}$ as to obtain the constant 
  \begin{equation}
    \lim_{E\downarrow 0}\; \frac{\ln \NDT(E)}{E^{-d/2}}=: -c_{*}(d,p)\,.
  \end{equation}
  An analogous statement is known from Thm.~{\normalfont 1.3} in
  \cite{BiKo01} for the case of \emph{site}-percolation graphs.
  Moreover, it is demonstrated in \cite{Ant95} that the bond- and the
  site-percolation cases have similar large-deviation properties.
\end{remarks}

The second main result of this subsection complements Theorem~\ref{lifshits} in the non-percolating
phase. 
\smallskip

\begin{thm}
 	\textup{\!(\cite{KiMu04}, Theorem 1.14)\,}
 \label{hove-subcritical}
  Assume $d \in\mathbb{N}$ and $p \in ]0,p_{c}[$. Then the
  integrated density of states of the Neumann Laplacians $(\DN_{G_{\omega}})_{\omega\in\Omega}$ on
  bond-percolation graphs in $\zd$ exhibits a \emph{Lifshits tail} with exponent $1/2$ at
  the lower spectral edge
  \begin{equation}
    \lim_{E\downarrow 0}\;\frac{\ln|\ln[ \NNn(E) - \NNn(0)]|}{\ln E} =
    -\frac{1}{2} \,,
  \end{equation}
  while that of the Dirichlet Laplacians $(\DD_{G_{\omega}})_{\omega\in\Omega}$ exhibits one at the upper
  spectral edge
  \begin{equation}
    \lim_{E \uparrow 4d}\;\frac{\ln|\ln [\ND^{-}(4d) - \ND(E)]|}{\ln (4d -E)} =  
    -\frac{1}{2} \,,
  \end{equation}
  where $\ND^{-}(4d) := \lim_{E\uparrow 4d} \ND(E) = 1- \NNn(0) $. 
\end{thm}

\begin{remarks}
	\item This theorem also follows from sandwich bounds analogous to those in Lemma~\ref{liflemma}. 
		We do not state them here but refer to Lemmas~2.7 and~2.9 in \cite{KiMu04} for details. 
		Using interlacing techniques, \cite{Sob05} establishes a better control on the constants in these bounds. For example, it was found that for all sufficiently small energies $E$
		\begin{equation}
 	\NNn(E) -\NNn(0) \le A E \exp\{ - \alpha_{+} E^{-1/2}\}
\end{equation}
with $\alpha_{+} := 4/[3\sqrt{3} \chi_{p}^{4}]$, where $\chi_{p}$ stands for the expected number of vertices in the cluster containing the origin and where the constant $A>0$ can also be made explicit.
	\item 
		\label{Nzero}
		The constant $\NNn(0)$ appearing in Theorem~\ref{hove-subcritical} is given by 
		    \begin{equation}
			 \NNn(0) = \lim_{\Lambda \uparrow\zd}
        \frac{\tr\nolimits^{\phantom{y}}_{\ell^{2} (\Lambda)} 1_{[0,\infty[} \bigl(
          -\DN_{G_{\omega}, \Lambda} \bigr) }{|\Lambda|} = \rho(p) + (1-p)^{2d}
    \end{equation}
		and equals the mean number density $ \rho(p)$ of clusters with at least two and at most finitely many 
		vertices, see e.g.\ Chap.~4 in \cite{Gri99}, plus the number density of isolated vertices. This follows from the fact that the operator $ 1_{[0,\infty[} (-\DN_{G_{\omega}, \Lambda} )$ is nothing but the
    projector onto the null space of the restriction $\DN_{G_{\omega}, \Lambda}$ of $\DN_{G_{\omega}}$ to $
    \ell^2(\Lambda)$. The dimensionality of this null space equals the number of finite clusters and 
    isolated vertices of $G_{\omega}$ in $\Lambda$, see Remark~1.5(iii) in \cite{KiMu04}.
	
  \item \label{linear-cluster}
  	The Lifshits tail for $\NNn$ at the lower spectral edge -- and hence
    the one for $\ND$ at the upper spectral edge -- is determined by the
    linear clusters of bond-percolation graphs. This explains why the
    associated Lifshits exponent $-1/2$ is not affected by the spatial
    dimension $d$.  Technically, this relies on a Cheeger inequality
    \cite{Col98} for the second-lowest Neumann eigenvalue of a connected
    graph, see also Prop.~2.2 in \cite{KiMu04}.
\end{remarks}

The third main result of this subsection is the counterpart of Theorem~\ref{hove-subcritical} in the percolating phase. 
\smallskip

\begin{thm}
  \label{hove}
 	\textup{\!(\cite{MuellerS-07}, Theorem 2.7)\;}
  Assume $d \in\mathbb{N}\setminus\{1\}$ and $p \in ]p_{c},1[$. Then the
  integrated density of states of the Neumann Laplacians $(\DN_{G_{\omega}})_{\omega\in\Omega}$ on
  bond-percolation graphs in $\zd$ exhibits a \emph{van Hove asymptotic} at
  the lower spectral edge
  \begin{equation}
    \label{lowerhove}
    \lim_{E\downarrow 0}\;\frac{\ln[ \NNn(E) - \NNn(0)]}{\ln E} =
    \frac{d}{2} \,,
  \end{equation}
  while that of the Dirichlet Laplacian $\DD_{G_{\omega}}$ exhibits one at the upper
  spectral edge
  \begin{equation}
    \label{upperhove}
    \lim_{E \uparrow 4d}\;\frac{\ln [\ND^{-}(4d) - \ND(E)]}{\ln (4d -E)} =  
    \frac{d}{2} .
  \end{equation}
\end{thm}

Similar to the two theorems above, Theorem~\ref{hove} also follows from upper and lower bounds and the symmetries \eqref{bcrel}.

\begin{lem}
  \label{hovelemma}
 	\textup{\;(\cite{MuellerS-07}, Lemma 4.1)\,}
  Assume $d\in\mathbb{N}\setminus\{1\}$ and $p\in]p_{c},1[$. Then there
  exist constants 
  $\varepsilon_{{N}}$, $C_{u}$, $C_{l} \in ]0,\infty[$ such
  that 
  \begin{equation}
    \label{hovebounds}
     C_{l} E^{d/2} \le \NNn(E) - \NNn(0) \le   C_{u} E^{d/2}
  \end{equation}
  holds for all $E\in ]0, \varepsilon_{{N}}[$.
\end{lem}

\begin{remarks}
\item Lemma~\ref{hovelemma} relies mainly on
  recent random-walk estimates \cite{MaRe04,Barlow-04,HeHo05} for the long-time decay of the
  heat kernel of $\DN_{G_{\omega}}$ on the infinite cluster. 
\item   There is also an additional Lifshits-tail
behaviour with exponent $1/2$ due to finite clusters as in Theorem~\ref{hove-subcritical}, but it is hidden under the dominating van Hove 
asymptotic of Theorem~\ref{hove}. Loosely speaking, Theorem~\ref{hove} is true because the
percolating cluster looks like the full regular lattice on very large length
scales (bigger than the correlation length) for $p>p_{c}$. On smaller scales
its structure is more like that of a jagged fractal. The Neumann Laplacian
does not care about these small-scale holes, however. All that is needed for
the van Hove asymptotic to be true is the existence of a suitable $d$-dimensional,
infinite grid. The adjacency and Dirichlet Laplacians though do care about those small-scale holes, as we infer from Theorem~\ref{lifshits}. 
\item In the physics literature the terminology van Hove ``singularity'' is also used for this kind of asymptotic. This refers to the fact that for odd dimensions $d$ derivatives seize to exist for high enough order.
\end{remarks}

The above three theorems cover all cases for $p$ and $X$ except the behaviour at the critical point $p=p_{c}$ of $\NNn$ at the lower spectral edge, respectively that of $\ND$ at the upper spectral edge. 
In dimension $d=2$ upper and lower power-law bounds have been obtained in \cite{Sob08}. However, the exponents differ so that the asymptotics is still an open problem; see also Remark~\ref{AO-conjecture} below for further properties at criticality.

%
%

\subsection{The regular infinite tree (Bethe lattice)}

In this subsection we report results from \cite{Rei09} on the asymptotics at spectral edges for the IDS of 
the family of Laplacians $(\Delta_{G_{\omega}}^X)_{\omega\in\Omega}$ on bond-percolation subgraphs of the $(\kappa+1)$-regular rooted infinite tree, a.k.a.\ Bethe lattice $\bbk$, where $\kappa\in\NN\setminus\{1\}$. Percolation on regular trees is well studied, see e.g.\ \cite{Per99}, and it turns out that the bond-percolation transition occurs sharply at the unique critical probability $p_{c} = \kappa^{-1}$. Here, sharpness of the phase transition is implied by, e.g., Theorem~\ref{sharp}, but it can also be verified by explicit computations. In contrast to percolation on the hypercubic lattice $\LL^d$, where the infinite cluster of the percolating phase is unique, there exist infinitely many percolating clusters simultaneously for $p>p_{c}$ on $\bbk$. 

The results on spectral asymptotics of the IDS are analogous in spirit to the ones of the previous subsection, but restricted to the non-percolating phase. However, as the Bethe lattice $\bbk$ exhibits an exponential volume growth of the ball $B(n)$ of radius $n$ about its root
$$
	V(n) = |B(n)| =  1 + (\kappa+1)\sum_{\nu=1}^n \kappa^{\nu -1} = 1 +  (\kappa^{n} -1)\; \frac{\kappa+1}{\kappa-1}, 
$$
cf.\ Figure~\ref{fig:2}, there will be natural differences. 

The next lemma determines the spectral edges of the operators under consideration. As a consequence of the exponential growth of the graph, and in contrast to the preceding subsection, the spectrum of the Laplacian on the Bethe lattice does not start at zero, neither does it extend up to twice the degree $2(\kappa+1)$.

\begin{lem}
	Let $\kappa\in\NN \setminus \{1\}$ and let $\Delta_{\bbk}$ be the Laplacian on the (full) Bethe lattice $\bbk$. Then
	$$
		\mathrm{spec}(\Delta_{\bbk}) = [E_{\kappa}^-, E_{\kappa}^+], \qquad \text{where~} E_{\kappa}^\pm := (\sqrt{\kappa} \pm 1)^2.
	$$
	Moreover, for $\PP$-almost every realisation $G_{\omega}$ of bond-percolation subgraphs of $\bbk$ we have 
	$$
		\mathrm{spec}(\Delta_{G_{\omega}}^N)  \subseteq [0, E_{\kappa}^+], \qquad
		\mathrm{spec}(\Delta_{G_{\omega}}^A)  =  [E_{\kappa}^-, E_{\kappa}^+], \qquad
		\mathrm{spec}(\Delta_{G_{\omega}}^D)  \subseteq [E_{\kappa}^-, 2(\kappa+1)].		
	$$
\end{lem}

\begin{remarks}
	\item
 	We believe that equality (and not only ``$\subseteq$'') holds for the statements involving the Neumann and the Dirichlet Laplacians, too. 
	\item
	Since the Bethe lattice is bipartite the above lemma reflects the symmetries \eqref{lap-symm}. 
	\item
	Almost-sure constancy of the spectra (i.e.\ independence of $\omega$) is again a consequence of ergodicity of the operators, see e.g.\ \cite{AcKl92} for a definition of the ergodic group action.
\end{remarks}

The ergodic group action on the Bethe lattice, which was referred to in the last remark above, is even transitive so that the IDS $N_{X}$ of the family $(\Delta_{G_{\omega}}^X)_{\omega\in\Omega}$ can be defined as in Definition~\ref{ids-def} with the fundamental cell $\mathcal{F}$ consisting of just the root. Clearly, $N_{X}$ will then obey the symmetry relations 
  \begin{equation}
    \label{bcrel-bethe}
    \begin{split}
      \NDT (E) &= 1 - \lim_{\varepsilon\uparrow 2(\kappa+1) -E} \NDT(\varepsilon)\,,
      \\
      N_{{D(N)}}(E) &= 1 - \lim_{\varepsilon\uparrow 2(\kappa+1) -E}
      N_{{N(D)}} (\varepsilon)
    \end{split}
  \end{equation}
  for all $E\in [0,2(\kappa+1)]$. 
  
Our first result concerns the asymptotic of $N_{N}$ at the lower edge, resp. of $N_{D}$ at the upper edge. Since these two spectral edges are unaffected by the exponential volume growth, it comes as no surprise that we find the same type of Lifshits tail as in the $\zd$-case.  

\begin{thm}
	\textup{\!(\cite{Rei09})\;}
 \label{bethe-linear}
  Assume $\kappa \in\mathbb{N} \setminus \{1\}$ and $p \in ]0,p_{c}[$. Then the
  integrated density of states of the Neumann Laplacians $(\DN_{G_{\omega}})_{\omega\in\Omega}$ on
  bond-percolation graphs in $\bbk$ exhibits a \emph{Lifshits tail} with exponent $1/2$ at
  the lower spectral edge
  \begin{equation}
    \lim_{E\downarrow 0}\;\frac{\ln|\ln[ \NNn(E) - \NNn(0)]|}{\ln E} =
    -\frac{1}{2} \,,
  \end{equation}
  while that of the Dirichlet Laplacian $\DD_{G_{\omega}}$ exhibits one at the upper
  spectral edge
  \begin{equation}
    \lim_{E \uparrow 2(\kappa +1)}\;\frac{\ln|\ln [\ND^{-}(2(\kappa+1)) - \ND(E)]|}{\ln (2(\kappa+1) -E)} =  
    -\frac{1}{2} \,,
  \end{equation}
  where $\ND^{-}(2(\kappa+1)) := \lim_{E\uparrow 2(\kappa+1)} \ND(E) = 1- \NNn(0) $. 
\end{thm}

\begin{remarks}
 \item These asymptotics are again determined by the linear clusters of bond-percolation graphs, cf.\ Remark~\ref{linear-cluster}. The interpretation of the reference value $\NNn(0)$ in terms of the cluster plus isolated vertex density is analogous to Remark~\ref{Nzero}.
\item In contrast to this Lifshits-tail behaviour in the subcritical phase, one expects $\NNn(E) - \NNn(0)$ to obey a power-law for small $E$ at the critical point $p_{c}$, caused by the finite critical clusters. This is not yet fully confirmed, but upper and lower algebraic bounds (with different exponents) follow from the random-walk estimates in \cite{Sob08}.
\item \label{AO-conjecture}
It should be noted that the power-law behaviour at $p_{c}$ mentioned in the previous remark is not the one referred to by the famous \emph{Alexander-Orbach conjecture} \cite{AlOr82}. The latter concerns the $E^{4/3}$-behaviour as $E\to 0$ of $\NNn(E)$ on the \emph{incipient infinite percolation cluster}. For the case of the Bethe lattice this asymptotic was proven in \cite{BaKu06}. (Here no subtraction of $\NNn(0)$ is necessary. Instead, one kind of conditions on the event that the origin belongs to an infinite cluster, see e.g.\ \cite{BoChKeSp01} for details of the definition.) The Alexander-Orbach conjecture says that the $E^{4/3}$-asymptotic should also hold for percolation in $\zd$ for every $d \ge 2$. Extensive numerical simulations indicate that this is not true in $d=2$ \cite{Gra99}. We refer to \cite{BuHa96} for a comprehensive discussion and further references from a Physics perspective. 
\end{remarks}

In order to reveal the characteristics of the Bethe lattice we now turn to the spectral edges $E_{\kappa}^{\pm}$.

\begin{thm}
	\textup{\!(\cite{Rei09})\;}
 	\label{bethe-stern}
  Assume $\kappa \in\mathbb{N} \setminus \{1\}$ and $p \in ]0,p_{c}[$. Then the
  integrated density of states of $(\DX_{G_{\omega}})_{\omega\in\Omega}$ on
  bond-percolation graphs in $\bbk$ exhibits a \emph{double-exponential tail} with exponent $1/2$ at
  the lower spectral edge
  \begin{equation}
  	\label{lower}
    \lim_{E\downarrow E_{\kappa}^-}\; \frac{\ln \big[\ln|\ln \NX(E)|\big]}{\ln (E- E_{\kappa}^-)} =
    -\frac{1}{2}  \qquad\quad\text{for $X=A,D$}
  \end{equation}
  and one at the upper spectral edge
  \begin{equation}
  \label{upper}
      \lim_{E\uparrow E_{\kappa}^+}\; \frac{\ln \big[\ln \big|\ln \big( 1- \NX(E)\big)\big|\big]}{\ln (E_{\kappa}^+ - E)}  =   -\frac{1}{2}  \qquad\quad\text{for $X=N,A$}.
  \end{equation}
\end{thm}

\begin{remarks}
\item The extremely fast decaying asymptotic of \eqref{lower} -- and similarly that of \eqref{upper} -- is determined by the lowest eigenvalues $ E \sim E_{\kappa}^- + R^{-2}$ of those clusters in the percolation graph which are large fully connected balls of radius $R$. Their volume is exponentially large in the radius, $V(R) \sim \e^{R} \sim \e^{(E- E_{\kappa}^-)^{-1/2}}$, and their probabilistic occurrence is exponentially small in the volume.
\item One would expect Theorem~\ref{bethe-stern} to be valid beyond the non-percolating phase. However, the region $p \ge p_{c}$ is still unexplored.
\item  A double-exponential tail as in \eqref{lower} will also be found in Theorem~\ref{double-exp-lamp} below. This concerns the lower spectral edge of the IDS for percolation on the Cayley graph of the lamplighter group, which is amenable. These double-exponential tails in two concrete situations should also be compared to the less precise last statement of Theorem~\ref{lif-cayley} below, which, however, holds for superpolynomially growing Cayley graphs of arbitrary, finitely generated, infinite, amenable groups.
\end{remarks}

%
%

\subsection{Equality and non-equality of Lifshits and van Hove exponents on amenable Cayley graphs}

... is almost the title of a paper by Antunovi\'c and Veseli\'c \cite{AntunovicV-09}. Here we record their main results. In our definition of the IDS in Subsection \ref{ids} above, two entirely different cases were treated. Let us first consider the deterministic case of the Laplacian on the full graph, denoted by $N_{\per}$. In our case of a quasi-transitive graph the geometry looks pretty regular; just like in the case of a lattice, the local geometry has the same local structure everywhere. Specializing to Cayley graphs this allows one to relate the asymptotic of $N_{\per}$ near $0$  to the volume growth $V(n)$ defined in \eqref{volume}. The latter is the same for the different Cayley graphs of the same group, see Theorem \ref{growth} above.
\begin{thm}
 Let $\Gamma$ be an infinite, finitely generated, amenable group, $G=G(\Gamma,S)$ a Cayley graph of $\Gamma$ and $N_{\per}$ the associated IDS. If $G$ has polynomial growth of order $d$, then
\begin{equation}
\label{star}
\lim_{E\downarrow 0}\frac{\ln N_{\per}(E)}{\ln E} =\frac{d}{2} .
\end{equation}
If $G$ has superpolynomial growth, then
$$
 \lim_{E\downarrow 0}\frac{\ln N_{\per}(E)}{\ln E} =\infty .
$$
\end{thm}
Proofs can be found in \cite{Varopoulos-87,Lueck-02}. Note that the limit appearing in \eqref{star} is exactly the zero order Novikov-Shubin invariant, where zero order refers to the fact that we deal with the Laplacian on $0$-forms, i.e., functions.

Next we turn to the asymptotic of the IDS $N_X$ of the corresponding percolation subgraphs. Again, Lifshits tails are found.

\begin{thm}
\label{lif-cayley}
\textup{\!(\cite{AntunovicV-09}, Theorem 6)\;}
Let $G=G(\Gamma,S)$ be the Cayley graph of an infinite, finitely generated, amenable group. Let $N_X$ be the IDS for the Laplacians $(\Delta_{G_{\omega}}^X)_{\omega\in\Omega}$ of percolation subgraphs of $G$ with boundary condition $X\in\{ A,D\}$ in the subcritical phase, i.e., for $p<p_c$. Then there is a constant  $a_p>0$ so that for all $E>0$ small enough
$$
N_D(E)\le N_A(E)\le \exp\left[ -\frac{a_p}{2}\; \tilde{V}\left( \frac{1}{2\sqrt{2} | S |}E^{-\frac12}-1\right)\right] ,
$$
where $\tilde{V}(t):=V(\lfloor t \rfloor)$, the volume $V(n)$ is given by \eqref{volume} and $\lfloor t \rfloor$ denotes the integer part of $t \in\RR$. 
If $G$ has polynomial growth of order $d$, then there are constants $\alpha_D^+,\alpha_D^->0$ so that for $E>0$ small enough
$$
 \exp\left[-\alpha_D^-E^{-\frac{d}{2}}\right]\le N_D(E)\le
N_A(E)\le \exp\left[-\alpha_D^+E^{-\frac{d}{2}}\right]  .
$$
If $G$ has superpolynomial growth, then
\begin{equation}
\label{superpol}
\lim_{E\downarrow 0}\frac{\ln |\ln N_D(E) |}{| \ln E|}=
\lim_{E\downarrow 0}\frac{\ln |\ln N_A(E) |}{| \ln E|}=\infty .  
\end{equation}
\end{thm}

Theorem~\ref{bethe-stern} and Theorem~\ref{double-exp-lamp} provide much more detailed information as compared to \eqref{superpol}, but only in two specific situations: the non-amenable free group with $n \ge 2$ generators and the amenable lamplighter group.    

The equality that is mentioned in the title of this subsection is now an easy consequence.

\begin{cor}
 In the situation of the preceding theorem the van Hove exponent and Lifshits exponents for $X\in\{ A,D\}$ coincide, i.e.,
$$
\lim_{E\downarrow 0}\frac{\ln |\ln N_D(E) |}{| \ln E|}=
\lim_{E\downarrow 0}\frac{\ln |\ln N_A(E) |}{| \ln E|}=\lim_{E\downarrow 0}\frac{\ln N_{\per}(E) }{ \ln E} .
$$
\end{cor}

Note that the asymptotic proved for $N_D$ and $N_A$ in the case of polynomially growing Cayley graphs is actually more precise than the double-log-limit that appears in the preceding corollary. For Cayley graphs with superpolynomial growth, a lower estimate is missing. However, for the lamplighter groups a more precise statement can be proven, see Theorem~\ref{lamp} below. 

The results of the previous section for the lattice case indicate that one should expect a different behaviour for the IDS $\NNn$ of the Neumann Laplacian at the lower spectral edge: it should be dominated by the linear clusters for $p< p_{c}$. This is indeed true.
 
 \begin{thm} 
 	\label{cayley-linear}
 	\textup{\!(\cite{AntunovicV-09}, Theorem 14)\;}
  In the situation of the previous theorem there exist constants $\alpha_N^+,\alpha_N^->0$ so that for all $E>0$ small enough
$$
 \exp\left[-\alpha_N^-E^{-\frac12}\right]\le N_N(E)-N_N(0)\le
	 \exp\left[-\alpha_N^+E^{-\frac12}\right] .
$$
 \end{thm}
The dimension $d$ is replaced by 1 in these estimates, since linear clusters are effectively one-dimensional and independent of the volume growth of $G$. This latter result remains true for quasi-transitive graphs with bounded vertex degree. 

As already announced, here are the more detailed estimates for the lamplighter group.

\begin{thm}
\label{lamp}
\textup{\!(\cite{AntunovicV-09}, Theorems 11 and 12)\;}
 Let $G$ be a Cayley graph of the lamplighter group $\ZZ_m\wr \ZZ$.
\begin{nummer}
 \item  There are constants $a_1^+,a_2^+>0$ so that for all $E>0$ small enough
$$
N_{\per}(E)\le a_1^+\exp\left[ -a_2^+E^{-\frac12}\right]  .
$$
 \item  For every $r>\frac12$ there are constants $a_{1,r}^-,a_{2-r}^->0$ so that for all $E>0$ small enough
$$
N_{\per}(E)\ge a_{1,r}^-\exp\left[ -a_{2,r}^-E^{-\frac{r}{2}}\right]  .
$$
 \item \label{double-exp-lamp} 
 For every $p<p_c$ there are constants $b_1,b_2,c_1,c_2>0$ so that for all $E>0$ small enough
$$
\exp\left[-c_1e^{c_2E^{-\frac12}}\right] \le N_D(E)\le N_A(E)\le
\exp\left[-b_1e^{b_2E^{-\frac12}}\right]  .
$$
\end{nummer}

\end{thm}

%
%

\subsection{Outlook: some further models}

To conclude, we briefly mention two other percolation graph models for which the Neumann Laplacian exhibits a Lifshits-tail behaviour with Lifshits exponent $\frac{1}{2}$ at the lower spectral edge $E=0$ in the non-percolating phase. As in the cases we discussed above, see Theorem~\ref{hove-subcritical} for the integer lattice, Theorem~\ref{bethe-linear} for the Bethe lattice and Theorem~\ref{cayley-linear} for amenable Cayley graphs, these Lifshits tails will also be caused by the dominant contribution of linear clusters. For this reason they occur quite universally, as long as the cluster-size distribution of percolation follows an exponential decay -- no matter how complicated the ``full'' graph $G$ may look like.    
This structure will not be seen by the linear clusters of percolation!

The first class of models \cite{MuellerR-07, MuellerR-10} consists of graphs $G$ which are embedded into $\RR^d$ (or, more generally, into a suitable locally compact, complete metric space) with some form of \emph{aperiodic order}. The celebrated Penrose tiling in $\RR^2$ constitutes a prime example. But one can consider rather general graphs whose vertices form a uniformly discrete set in $\RR^d$ and whose edges do not extend over arbitrarily long distances. Amazingly, the main point that needs to be dealt with to establish Lifshits tails for such models concerns the definition of the IDS. In contrast to the definition in \eqref{ids-random}, one cannot expect to benefit from a quasi-transitive group action on $G$ with a finite fundamental cell in this aperiodic situation. The way out is to consider the \emph{hull} of the graph $G$, that is the set of all $\RR^d$-translates of $G$, closed in a suitable topology which renders the hull a compact dynamical system. As such it carries at least one $\RR^d$-ergodic probability measure $\mu$, and the expectation in \eqref{ids-random} will be replaced by a two-stage expectation: one with respect to $\mu$ over all graphs $G'$ in the hull of $G$, and inside of it, for each graph $G'$, the expectation $\EE_{p}^{(G')}$ over all realisations of percolation subgraphs of $G'$. The interested reader is referred to \cite{ Len09, MuellerR-10} for more details.      

The second model, \emph{Erd\H{o}s-R\'enyi random graphs} \cite{ErRe60, Bol01}, has a combinatorial background. There we consider bond percolation on the complete graph $K_{n}$ over $n$ vertices with bond probability $p:= c/n$. The $n$-independent parameter $c>0$ corresponds to twice the expected number density of bonds, if $n$ is large. This is sometimes referred to as the \emph{(very) sparse case}. For $c\in ]0,1[$, the fraction of vertices belonging to tree clusters tends to $1$ as $n\to\infty$, and the limiting cluster-size distribution decays exponentially. In this model the IDS is defined by
$$
N_{N}(E) := \lim_{n\to\infty} \EE_{c/n}^{(K_{n})} \big[ \langle \delta_{1}, 1_{]-\infty, E]}(\Delta_{G_{\omega}}^N) \delta_{1} \rangle \big],
$$
and it exhibits a Lifshits tail at the lower spectral edge $E=0$ with exponent $1/2$ \cite{KhKi06}.
%
%


\subsection*{Acknowledgment}
Many thanks to the organisers of the Alp-Workshop at St.\ Kathrein for the kind invitation and the splendid hospitality extended to us there.
\end{document}